# Inelastic Cotunneling Resonances in the Coulomb-Blockade Transport in Donor-Atom Transistors

Pooja Yadav[1], Soumya Chakraborty[1], Daniel Moraru[2], Arup Samanta[1,3]

[1]Department of Physics, Indian Institute of Technology Roorkee, Roorkee-247667, Uttarakhand, India
[2]Research Institute of Electronics, Shizuoka University, 3-5-1 Johoku, Naka-ku, Hamamatsu 432-8011, Japan
[3]Centre of Nanotechnology, Indian Institute of Technology Roorkee, Roorkee-247667, Uttarakhand, India

We report finite-bias characteristics of electrical transport through phosphorus donors in silicon nanoscale transistors, in which we observe inelastic-cotunneling current in the Coulomb blockade region. The cotunneling current appears like a resonant-tunneling current peak emerging from the excited state at the crossover between blockade and non-blockade regions. These cotunneling features are unique, since the inelastic-cotunneling currents have so far been reported either as a broader hump or as a continuous increment of current. This finding is ascribed purely due to excitation-related inelastic cotunneling involving the ground and excited states. Theoretical calculations were performed for a two-level quantum dot, supporting our experimental observation.

## Introduction

Continuous downscaling of transistors has enhanced the role of discrete dopant atoms in the transport through nanoscale channels. Long coherence times for spin [1,2] and charge [3] of electrons in dopants (such as phosphorus (P) donors) are key advantages for one of the quantum computing architectures. With donors now becoming an active element of nanoscale transistors, their location and number play a vital role in the performance of the devices. Donor atoms mimic quantum dots (QDs) with discrete energy levels. Single-electron tunneling (SET) through such donor-QDs has been already achieved and reported in nanoscale field-effect transistors (FETs) [4–11]. In such nanoscale FETs, with a small number of donors in the channel region, Coulomb-blockade tunneling transport occurs at low temperature through the ground state of the donor for a first and, under certain conditions, a second electron, namely by transitions that can be labeled as $D^+/D^0$ and $D^0/D^-$, respectively [4,5]. The characteristic charging and binding energies of the donor-based devices can be modified by quantum confinement [12], dielectric confinement [13-15] and by coupling of multiple donors [16–18]. High-temperature SET transport was reported at around 100 K for single-donor QDs embedded in stub-shaped channels by using the dielectric confinement effect [14] and even at more elevated temperatures (~150 K) for multiple-donor QDs formed by selective doping [18]. It is expected that at low bias transport is dominated by the ground state, while at higher bias excited states also contribute to the current.

In single-electron tunneling through QDs, Coulomb blockade suppresses the sequential transport of electrons, but some electrons may escape the blockade and tunnel via the Coulomb energy gap between different charge states. These tunneling processes inside the blockade region arise due to quantum fluctuations of electric charge on the QD, resulting in small "leakage" current flowing within the non-conducting region. This current is known as cotunneling current [19], since it involves simultaneous tunneling of multiple electrons. Cotunneling events are classified into two categories based on the use of different transport channels. In elastic cotunneling, electrons cross the two barriers using the ground state of the QD, leaving the QD in the ground state. On the other hand, in the inelastic cotunneling, different states are involved to cross the barriers and the QD can be left in the excited state. Although single-electron tunneling is forbidden by the Coulomb blockade condition, electrons are still allowed to spend some time in the forbidden state because of the uncertainty principle. The estimated time for the process is bound by $\frac{\hbar}{\Delta E}$, where $\Delta E$ is the energy difference between successive states [20].

Extensive theoretical [21−23] and experimental works on cotunneling transport have been performed for semiconductor QDs [24,25], donors [11,26] and molecules [27], where multiple levels contribute to the cotunneling features. One theoretical model for cotunneling is provided by the macroscopic quantum tunneling (Q-mqt) theory [28]. A theoretical model of inelastic cotunneling using discrete energy levels has been also proposed to explain experimental results [21,29]. In general, inelastic cotunneling current starts from the point where the excited state is practically aligned with the Fermi level of the reservoirs, filling the whole blockade region above the excited state. However, we report inelastic-cotunneling current in a donor-atom transistor with features similar to resonant current peaks. Here, we experimentally and theoretically study such inelastic-cotunneling current in donor-atom transistors.

## Experimental Results:

We fabricated FETs in a silicon-on-insulator (SOI) wafer, having a buried oxide (BOX) layer of thickness of 150 nm, as presented in Figs. 1(a)-(d). All process steps for the device fabrication were performed in a clean room environment, using CMOS-compatible techniques. The final channel widths of the transistor channels are in the range of 10–100 nm. A scanning electron microscope (SEM) image of one of the devices (before gate oxide formation) is presented in Fig. 1(b). Doping of P-donors was carried out by thermal diffusion in the entire SOI layer, resulting in a donor concentration of $N_\mathrm{D} \approx 10^{18}$ cm$^{-3}$ (same in the channel, source and drain). A typical donor distribution is schematically presented in Fig. 1(c). Gate oxide was grown by dry thermal oxidation, before the formation of the Al gate, source and drain electrodes. The final thickness of the SOI layer of the FETs is estimated to be ≈ 5 nm and the gate-oxide thickness is $t_\mathrm{ox} \approx 10$ nm, as seen also from the cross-sectional transmission electron microscope (TEM) image shown in Fig. 1(d). Some of the electrical characteristics for similar transistors have been reported earlier [14, 30, 31].

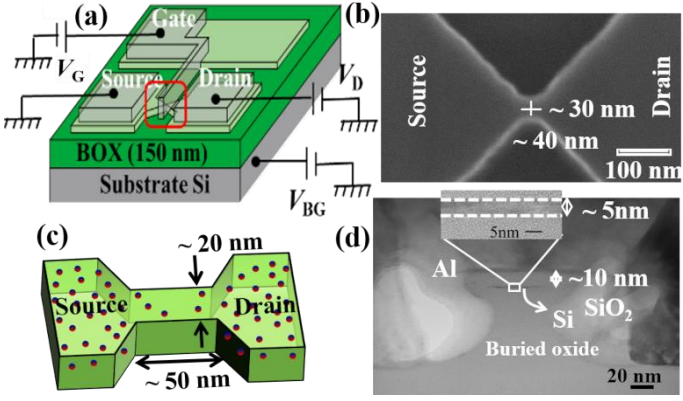

**Figure 1:** (Color online) (a) Schematic configuration of an SOI-FET. (b) Scanning electron microscope image of the channel region in one of the SOI-FETs before the gate-oxide formation. (c) Schematic distribution of donors in the channel region. (d) Cross-sectional image of an SOI-FET measured by transmission electron microscope.

In this report, we present electrical characteristic of one such device at low temperature. The estimated dimensions of the channel region of this device are ≈ 50 nm, ≈ 20 nm, and ≈ 5 nm for length, width, and thickness, respectively. The estimated number of P-donors in the channel region is approximately 5. Electrical measurements of the device were performed in high vacuum and at low temperature, $T$ = 5.5 K. For these measurements, the bias voltage ($V_{DS}$) was applied to the drain electrode with respect to the grounded source electrode, while the gate voltage ($V_G$) was used as a variable parameter.

Drain current ($|I_{DS}|$) vs gate voltage ($V_G$) characteristics of the device were measured for different $V_{DS}$ values. The $|I_{DS}|$-$V_G$ characteristics for $V_{DS}$ = -30 mV are presented in Fig. 2(a). Several aperiodic and dissimilar isolated current peaks (marked as $P_1$, $P_2$ and $P_3$) can be observed at low $V_G$'s and can be ascribed to single-electron tunneling via different P-donors (as will be argued later), before the onset of a larger current due to FET transport over the barrier. Figure 2(b) shows the stability diagram (displaying the conductance, $dI_{DS}/dV_{DS}$, in the plane of $V_G$-$V_{DS}$) for the same device. Irregularities in the shape of the Coulomb diamonds suggest that transport occurs through multiple QDs, which is consistent with initial expectations because several P-donors are in the channel. Coulomb blockade works to suppress the first-order sequential tunneling current within the diamond-shaped regions. The first-order tunneling takes place basically when the ground state of a P-donor QD is aligned by $V_G$ application with the chemical potential (Fermi level) of source and drain reservoirs. The boundaries of the Coulomb diamonds are roughly drawn as guides for the eye.

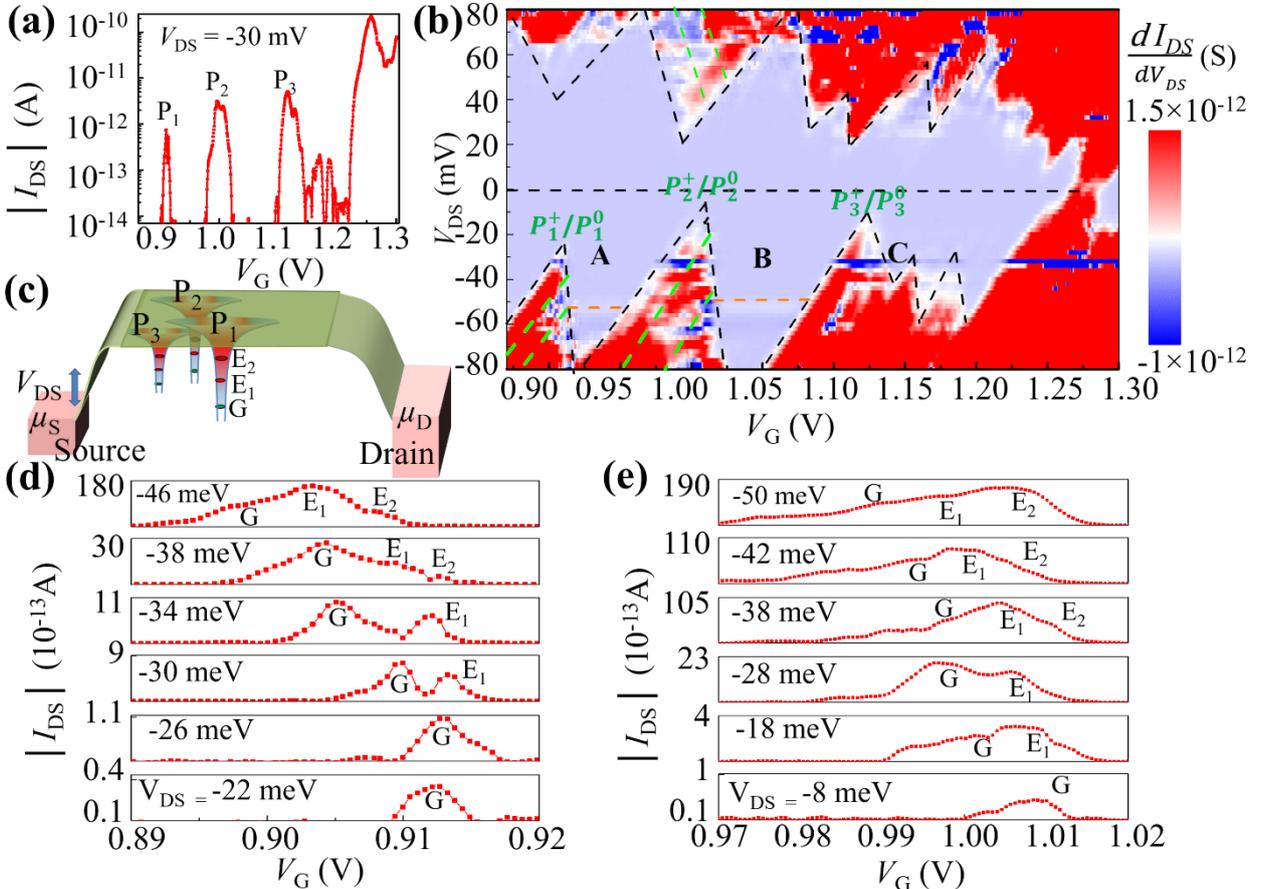

**Figure 2:** (Color online) (a) $I_{DS}$-$V_G$ characteristics of the donor-atom transistor made with an SOI-FET configuration. (b) Stability diagram (plot of $dI_{DS}/dV_{DS}$ in the plane of $V_G$-$V_{DS}$) for the transport characteristics of the same device. (c) Schematic diagram for the distribution of donors along with their potential in the channel region of the device. (d)-(e) $I_{DS}$-$V_G$ characteristics measured at different bias voltages to show the signatures of transport through ground and excited states for the $P_1$ and $P_2$ donors, respectively.

The charge transition through the donor $P_1$, i.e., $P_1^+/P_1^0$, followed by the Coulomb-blockade region A, are marked in Fig. 2(b). We have calculated the lever-arm factor $C_G/(C_G + C_S + C_D)$, $\alpha \approx 0.82$, and estimated the position of the $P_1$ donor with respect to the source reservoir as $L_S \approx 0.47 \times L$, where $L$ is the channel length and $C_G$, $C_S$ and $C_D$ are the gate, source and drain capacitances, respectively, for this QD. Similarly, we identified the charge transitions $P_2^+/P_2^0$ and $P_3^+/P_3^0$ through the donor $P_2$ (followed by the Coulomb-blockade region B) and through the donor $P_3$ (followed by the Coulomb-blockade region C), respectively. The lever-arm factors for the $P_2$ and $P_3$ donors are approximately 0.70 and 0.75, respectively, while their locations relative to the source reservoir were estimated as $0.32 \times L$ and $0.22 \times L$, respectively. The different $\alpha$ values and different estimated locations of the QDs, strongly suggest that these conductance features arise from different donors. The schematic potential distribution for the donor-QDs in the channel region is presented in Fig. 2(c). It should be noted that we observed open diamonds in our device, which is most likely due to high-resistance tunnel barriers, with the tunnel resistance being bias dependent, as we reported in our previous work [32].

We also observed multi-level sequential transport in each of the non-blockade region of the stability diagram. At higher bias, the onset of the first-order tunneling involves both ground and excited states. This process can be exploited as a spectroscopic tool to determine the discrete energy spectrum of the donors [4-6]. We mainly observed two excited states for each donor. The excited-state current features are highlighted by the green dotted lines in the stability diagram. We also investigated such transport features from the $I_{DS}$-$V_G$ characteristics, as presented in Figs. 2(d) and 2(e) for the donors $P_1$ and $P_2$, respectively. In this data, we observed a single, smooth current peak for the $P_1$ donor up to $V_{DS} = -26$ mV, which suggests that only the ground state (labeled $G$) is accessible. At the higher bias voltage, the first excited state ($E_1$) and the second excited state ($E_2$) are clearly manifested. For the $P_2$ donor, as marked in Fig 2(e), we also observed transport through the ground state at low bias and through the excited states, $E_1$ and $E_2$, at higher bias. Considering the lever-arm factors estimated above, we also estimated $E_1$ and $E_2$ with respect to the ground states. For the $P_1$ donor, $E_1$ and $E_2$ are $\approx 4$ meV and $\approx 8.5$ meV, respectively, while for the $P_2$ donor, $E_1 \approx 4.5$ meV and $E_2 \approx 9$ meV. These values for the excited states are smaller as compared to the bulk-Si donor spectrum [33,34], but are consistent with donors located near the Si/SiO$_2$ interface [6], as is the case in our devices.

Now, we will discuss the main result of the present investigation, i.e., the resonant-like leakage current in the Coulomb blockade region of the stability diagram, observed along the $V_G$ axis, as marked by the orange dotted lines in the regions A and B. In principle, no sequential transport should occur in the Coulomb blockade region according to the orthodox theory of Coulomb blockade. However, a new transport channel may open up when the tunnel barriers are relatively low. Here, we are interested in the Coulomb blockade region where higher-order (mainly second-order) tunneling processes dominate. These features are expected to exist on both sides of the Coulomb blockade region (for $V_{DS}$ both positive and negative). However, we observed distorted Coulomb diamonds in the positive-$V_{DS}$ side, which is likely due to a more complex array of QDs for such bias polarity. Hence, we will mainly analyze the negative-bias region of the stability diagram for further discussion.

From now, we will more closely analyze the unconventional conductance features observed parallel to the $V_G$ axis within the Coulomb blockade region. For that purpose, the regions A and B in the stability diagram shown in Fig. 2(b) are zoomed in and re-plotted in Figs. 3(a) and 3(b), respectively. We can observe from Fig. 3(a) that a sharp resonant-like differential conductance line connects two conducting regions, as marked by the orange dotted line. This line is originated at the second excited state of the $P_1$ donor. Similar observation is made for the $P_2$ donor, as presented in Fig. 3(b). To emphasize more on this part, we plotted $dI_{DS}/dV_{DS}$ and current for $P_1$ and $P_2$ donors along the orange dotted lines in Figs. 3(c) and 3(d), respectively. We observed a sharp conductance peak inside the blockade region when the bias voltage is such that the Fermi level is aligned with the level $E_2$ for each case.

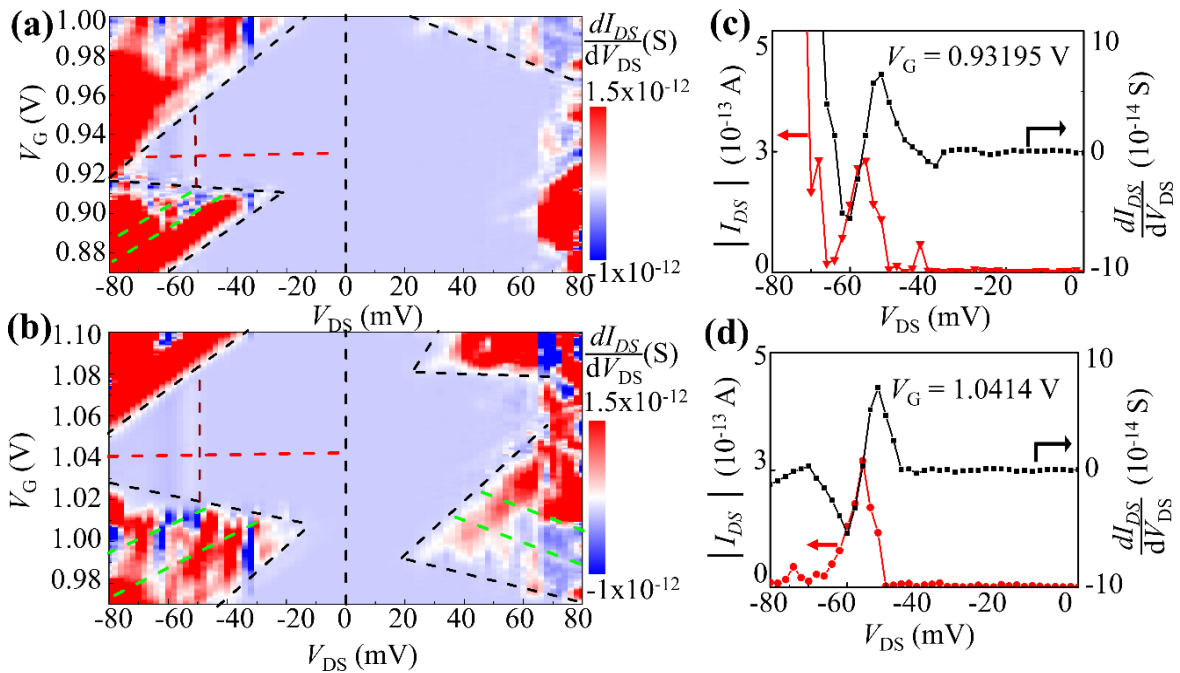

**Figure 3:** (Color online) (a)-(b) Conducting and blockade regions corresponding to transport through the donors $P_1$ and $P_2$, respectively. (c)-(d) Current and differential conductance data, taken along the dashed lines for the $P_1$ and $P_2$ donors, respectively.

We propose that these current features, observed inside the blockade regions, can be ascribed to the inelastic cotunneling current associated with the $E_2$ state of the donors. It is important to note that, in other reports, one frequently observes a broad hump [24,25] or purely continuously increasing high current [19], rather than a sharp peak as signatures of cotunneling current. To explain the observed unconventional inelastic-cotunneling current features, we have theoretically calculated the inelastic cotunneling in the next section.

### Theoretical Formalism and Discussion:

Let us quantify the inelastic cotunneling in a SET device. Inelastic cotunneling is due to higher-order transitions occurring in the Coulomb blockade region. To estimate the inelastic cotunneling involving the ground state and the excited state of the QD in the SET device, we have used second-order transition including a mechanism equivalent to electron-hole excitation as a base for the calculation model.

The Anderson Hamiltonian of a QD having two single-particle energy levels $\epsilon_1$ and $\epsilon_2$ connected to electron reservoirs (source and drain), excluding the spin contribution, is described as [35]:

$$H = H_{Dot} + H_{S(D)} + H_T \quad (1)$$

where the Hamiltonian for coupling of the QD to source and drain reservoirs is described by $H_T$, while the Hamiltonians for an ideal (isolated) QD and for the source (drain) reservoirs are described by $H_{Dot}$ and $H_{S(D)}$, respectively. Here:

$$H_{S(D)} = \sum_{i \in S(D)} \varepsilon_i c_i^\dagger c_i$$

$$H_{Dot} = \sum_{j(=1,2)} \varepsilon_j b_j^\dagger b_j + E_c b_1^\dagger b_1 b_2^\dagger b_2$$

$$H_T = \sum_{i \in S(D), j(=1,2)} (U_{ij} c_i^\dagger b_j + \text{H.c}) \quad (2)$$

where the energy of the quasiparticles in the reservoirs is $\varepsilon_i$ and $\varepsilon_j$ ($j=1,2$) corresponds to the two energy states of the QD considered here; $E_C$ indicates the coulombic interaction among electrons in the QD, with the transition rate following the Fermi golden rule between the QD's energy level $\epsilon_j$ and the reservoir:

$$\Gamma_{i \in S(D)}^j (\varepsilon_j) = \frac{2\pi}{\hbar} \sum_{i \in S(D)} |<i|M|j>|^2 \delta(\epsilon_i - \epsilon_j) \quad (3)$$

Since the tunneling processes rely upon the bias condition, the initial and final states may vary depending upon the bias. The total tunneling rates from the initial state $i$ to the final states $f$ through intermediate virtual states $v$ depend upon the transition matrix $M_{if} = \sum_{j(=1,2)} \frac{H_T^{(ij)} H_T^{(jf)}}{E_i - E_j}$ and are expressed as [20]:

$$\gamma_{i \in S(D)} = \frac{2\pi}{\hbar} \sum_f |<i|M|f>|^2 \delta(\epsilon_i - \epsilon_f), \quad (4)$$

$\gamma_S(V_{DS})$ or $\gamma_D(V_{DS})$ are transition rates depending upon the starting point of the initial electron transition, i.e., either from source or drain. $\gamma_S(V_{DS})$ can be now expressed as [22]:

$$\gamma_S(V_{DS}) = \frac{\hbar \Gamma_S \Gamma_D}{2\pi} \int d\epsilon d\epsilon' f(\epsilon)[1 - f(\epsilon')]$$
$$\left[ <n_1> \left( \frac{1}{\epsilon_2 - \epsilon + E_S} + \frac{1}{\epsilon' - \epsilon_1 + E_D} \right)^2 \delta(\epsilon' - \epsilon_1 + \epsilon_2 - \epsilon - eV_{DS}) \right.$$
$$\left. + <n_2> \left( \frac{1}{\epsilon_1 - \epsilon + E_S} + \frac{1}{\epsilon' - \epsilon_2 + E_D} \right)^2 \delta(\epsilon' + \epsilon_1 - \epsilon_2 - \epsilon - eV_{DS}) \right]$$

(5)

where $\Gamma_S$ ($\Gamma_D$) is the transition rate between an electron energy level $\epsilon_j$ and source (drain) reservoir, while $f(\varepsilon) = \frac{1}{1 + e^{\frac{\varepsilon}{kT}}}$ is the Fermi-Dirac distribution function defined for the reservoirs.

$$<n_1> = \frac{1}{1 + e^{-\frac{\epsilon_2 - \epsilon_1}{kT}}}$$

$$<n_2> = \frac{e^{-\frac{\epsilon_2 - \epsilon_1}{kT}}}{1 + e^{-\frac{\epsilon_2 - \epsilon_1}{kT}}}$$

$$E_S = nE_c - \mu_S \text{ and } E_D = \mu_D - (n-1)E_c$$

Here, $E_S$ and $E_D$ are charging energies of virtual energy levels with an extra electron (or hole, respectively); $\mu_s$ ($\mu_d$) are the chemical potentials of the source (drain) reservoir, satisfying the equation $\mu_s - \mu_d = eV_{DS}$ and $n$ signifies the number of electrons in the QD.

The first term in the summation in Eq. 5 is the excitation process and the second term is the relaxation process. If the excitation is the main dominating process, Eq. 5 can be further reduced (following the ref. [22]) to the final equation as:

$$\gamma_S(V_{DS}) = \frac{\hbar \Gamma_S \Gamma_D}{2\pi} \left( \frac{1}{E_S + \frac{\Delta}{2}} + \frac{1}{E_D + eV_{DS} - \frac{\Delta}{2}} \right)^2 (\Delta - eV_{DS}) e^{-\frac{\Delta - eV_{DS}}{kT}}$$

(6)

Here, we consider $\varepsilon_1 = 0$ and $\varepsilon_2 = \Delta$.

We can now calculate the inelastic cotunneling current using the formula as given below:

$$I_{inelastic} = \gamma_S(V_{DS}) - \gamma_D(V_{DS})$$

and

$$\gamma_D(V_{DS}) = \gamma_S(-V_{DS}) \quad (7)$$

Hence, the final expression for the inelastic cotunneling current is:

$$I^{inelastic} = \frac{e \hbar \Gamma_S \Gamma_D}{2\pi} \left[ \left( \frac{1}{E_S + \frac{\Delta}{2}} + \frac{1}{E_D + eV_{DS} - \frac{\Delta}{2}} \right)^2 (\Delta - eV_{DS}) e^{-\frac{\Delta - eV_{DS}}{kT}} \right.$$

$$-\left(\frac{1}{E_S + \frac{\Delta}{2}} + \frac{1}{E_D - eV_{DS} - \frac{\Delta}{2}}\right)^2 (\Delta + eV_{DS}) e^{-\frac{\Delta + eV_{DS}}{kT}}\bigg]$$

(8)

Equation 8 can basically interpret the observed inelastic cotunneling current in our experimental results. For that purpose, we have numerically calculated the stability diagram for a single-electron transistor having two energy levels in the quantum dot by using the modified rate equation [36] and the inelastic cotunneling contribution obtained from this analysis. To perform this calculation, we have used the following device parameters: the total quantum dot capacitance $C_\Sigma = 5 \times 10^{-18}$ F, temperature $T = 5$ K and a level spacing $\Delta = 10$ meV, comparable to the values estimated from the experimental results.

The numerical calculation is presented as a contour plot in Fig. 4(a). Current starts in the blockade region from $|V_{DS}| = 10$ mV, then takes a sudden dip and rises again at the Coulomb-diamond boundary. The inelastic cotunneling current present in the blockade area is clearly visible by a greenish current line aligned with the $V_G$ axis. This current can be seen originating from the excited state, observable as the current step in the non-blockade region. The cotunneling current peak is represented in Fig. 4(b) represents the same results, but visualized in the $|I_{DS}|$-$V_{DS}$ characteristics in the blockade region. The current peak appearing at the excitation gap $\Delta$ is a result of electrons being excited from source to $\varepsilon_2$ energy level (and hole being excited from $\varepsilon_1$ energy level to drain reservoir). The present experimental results from the donor-atom SOI-FET exhibit similar blockade current features. The sharp current peak positioned at bias voltage corresponding to the excited state resembles resonant-like characteristics and such a feature is also clearly observed in our device data. As mentioned previously, in most experiments reported so far, the observed cotunneling features are either seen as continuously increasing current or a broader hump starting from the excited sates, which is contrary to our observation. The observation of such conventional features is ascribed due to contribution from excitation related inelastic cotunneling along with the additional contribution from the relaxation processes and sequential tunneling in the blockade region [29, 37].

Combining all the tunneling regimes, the specific Coulomb-blockade behavior appearing in such devices can be identified and sketched in Fig. 4(c), where it is shown that only the excitation process is dominating. In the conducting portion, the transport is taking place involving two states, while in the blockade region, a virtual excited state and the ground state are allowing the transport. The arrows point towards illustrations of different regimes of the transport. It can be thus stated that the theoretical current features are purely excitation-related inelastic cotunneling in nature, without considering any relaxation and sequential processes [21,29].

The experimental observation of the inelastic cotunneling, as presented in Fig. 2 and Fig. 3, is, therefore, in agreement with the theoretical results and is a proof of leakage current through excited states despite the conditions for

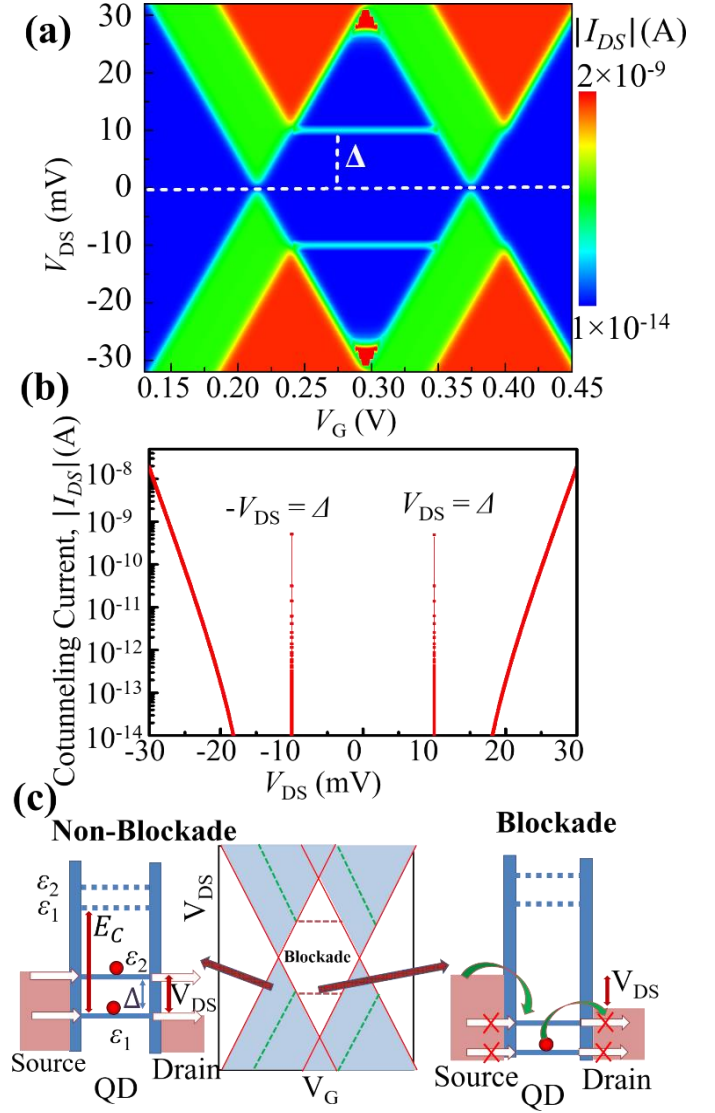

**Figure 4:** (Color online) (a) Contour plot of the sequential and inelastic cotunneling current of a theoretical device having ground and one excited (10 meV) states, where excited state and cotunneling features are clearly marked at $V_{DS} = \Delta = 10\ mV$. (b) The plot of inelastic cotunneling current with $V_{DS}$, where the sharp conductance peaks occurred at $|V_{DS}| = 10\ mV$. (c) Mechanism of electron transport through the SET device in the different regions of the stability diagram is schematic presented with different transport diagrams.

Coulomb blockade being respected. However, we observed inelastic cotunneling current through the second excited states of the donor. It is important to note here that the inelastic cotunneling rate is always bounded by the tunnel and quantum resistances as: $\gamma_{cot} = \gamma_{seq}\left(\frac{R_Q}{R_t}\right)$, where $R_Q$ is the quantum resistance and $R_t$ is the tunnel resistance. To observe the inelastic cotunneling, a suitable barrier condition is highly required. In our device, we do not observe any cotunneling current through the first excited state most likely due to the high barrier offered by this first excited state of the donor, while the observation of cotunneling through the second excited state is because of the lower resistance offered at this energy level. In our donor-atom device, the second excited state provides a more transmissive path for current leakage, leading to the observation of resonant-like inelastic cotunneling features purely due to the excitation-related transitions in the Coulomb blockade region.

## Conclusion

In summary, we showed Coulomb-blockade transport through a few phosphorus donors in a nano-channel SOI-FET at low temperature (5.5 K), and evidenced the importance of the discrete energy spectrum of the P-donors. Most importantly, we observed a sharp inelastic cotunneling current from the second excited state of the donors, appearing as resonant tunneling within the blockade region. We have theoretically investigated such features using second-order transitions and demonstrated that the observed current in the blockade region is purely due to excitation-related inelastic cotunneling involving the ground state and excited state, without any contribution from sequential and relaxation processes.

Since cotunneling has been identified as a possible source of uncertainty in the operation of single-electron tunneling devices, this phenomenon limits the performance of such devices for practical uses. Hence, the fundamental understanding provided by these results will help us to design and fabricate practical devices for developing the donor-based quantum architectures.

## Acknowledgement

We appreciate critical comments and discussions with M. Tabe for design of devices and analysis of the result. The authors thank T. Mizuno and R. Nakamura for contributions to the device fabrication. The authors acknowledge M. K. Sharma and P. Sudha for fruitful discussions. P.Y. and S.C. acknowledge the Ministry of Education for fellowship. The work is partially supported by DST-SERB (Project no: ECR/2017/001050) and IIT Roorkee (Project no: FIG-100778-PHY), India.

## References:

1. B. E. Kane. Nature **393** 6681 (1998).
2. A. Morello, C. C. Escott, H. Huebl, L.H.W. Van Beveren, L. C. L. Hollenberg, D.N. Jamieson, A.S. Dzurak, and R. G. Clark, Phys. Rev. B **80,** 081307 (2009).
3. L. C. L. Hollenberg, A. S. Dzurak, C. Wellard, A. R. Hamilton, D. Reilly, G. J. Milburn and R. G. Clark, Phys. Rev. B **69,** 113301 (2004).
4. H. Sellier, G. P. Lansbergen, J. Caro, S. Rogge, N. Collaert, I. Ferain, M. Jurczak, and S. Biesemans, Phys. Rev. Lett. **97,** 206805 (2006).
5. M. Fuechsle, J. A. Miwa, S. Mahapatra, H. Ryu, S. Lee, O. Warschkow, L. C. Hollenberg, G. Klimeck, and M. Y. Simmons, Nat. Nanotechnol. **7,** 242 (2012).
6. G. P. Lansbergen, R. Rahman, C. J. Wellard, I. Woo, J. Caro, N. Collaert, S. Biesemans, G. Klimeck, L. C. L. Hollenberg, and S. Rogge, Nat. Phys. **4** 8 (2008).
7. Y. Ono, K. Nishiguchi, A. Fujiwara, H. Yamaguchi, H. Inokawa, and Y. Takahashi, Appl. Phys. Lett. **90,** 102106 (2007).
8. D. Moraru, A. Samanta, K. Tyszka, M. Muruganathan, T. Mizuno, R. Jablonski, H. Mizuta, and M. Tabe, Nanoscale Res. Lett. **10,** 372 (2015).
9. M. Tabe, D. Moraru, M. Ligowski, M. Anwar, R. Jablonski, Y. Ono, and T. Mizuno, Phys. Rev. Lett. **105,** 016803 (2010).
10. E. Prati, M. Hori, F. Guagliardo, G. Ferrari, and T. Shinada, Nat. Nanotechnol. **7** 443-447 (2012).
11. M. Pierre, R. Wacquez, X. Jehl, M. Sanquer, M. Vinet, and O. Cueto. Nat. Nanotechnol. **5,** 2 (2010).
12. M. Diarra, C. Delerue, Y. M. Niquet, and G. Allan, J. Appl. Phys. **103,** 073703 (2008).
13. M. Diarra, Y. M. Niquet, C. Delerue, and G. Allan, Phys. Rev. B, **75,** 045301 (2007).
14. E. Hamid, D. Moraru, Y. Kuzuya, T. Mizuno, H. Mizuta, and M. Tabe, Phys. Rev. B, **87,** 085420 (2013).
15. M. T. Björk, H. Schmid, J. Knoch, H. Riel and W. Riess, Nat. Nanotechnol. **4**, 103 (2009).
16. D. Moraru, A. Samanta, L. T. Anh, T. Mizuno, H. Mizuta, and M. Tabe, Sci. Rep., **4,** 6219 (2014).
17. B. Weber, Y. M. Tan, S. Mahapatra, T. F. Watson, H. Ryu, R. Rahman, L. C. Hollenberg, G. Klimeck, and M. Y. Simmons, Nat. Nanotechnol. **9,** 430 (2014).
18. A. Samanta, M. Muruganathan, M. Hori, Y. Ono, H. Mizuta, M. Tabe, and D. Moraru, Appl. Phys. Lett. **110,** 9 (2017).
19. D. V. Averin and Y. V. Nazarov, Phys. Rev. Lett. **65,** 19 (1990).
20. Single Charge Tunneling H Grabert Michel H. Devoret, Proc. a NATO Adv. Study Inst. Chapter 6, 217-223, (1992).
21. V. N. Golovach, and D. Loss, Phy. Rev. B **69,** 245327 (2004).
22. K. Kang and B. I. Min, Phys. Rev. B **55,** 15412 (1997).
23. F. Cheng. and W. Sheng, J. Appl. Phys. **108,** 043701 (2010).
24. S. De Franceschi, S. Sasaki, J. M. Elzerman, W. G. Van Der Wiel, S. Tarucha, and L. P. Kouwenhoven, Phys. Rev. Lett. **86,** 878 (2001).
25. R. Schleser, T. Ihn, E. Ruh, K. Ensslin, M. Tews, D. Pfannkuche, D. C. Driscoll, and A. C. Gossard, Phys. Rev. Lett. **94,** 206805 (2005).
26. A. Crippa, M. L. V. Tagliaferri, D. Rotta, M. De Michielis, G. Mazzeo, M. Fanciulli, R. Wacquez, M. Vinet, and E. Prati, Phys. Rev. B **92,** 035424 (2015).
27. G. Begemann, S. Koller, M. Grifoni, and J. Paaske. Phys. Rev. B **82,** 045316 (2010).
28. D.V. Averin, and A. A. Odintsov, Phys. Lett. A **140,** 5 (1989).
29. Y. Nazarov, and Y. Blanter, Quantum Transport: Introduction to Nanoscience. Cambridge: Cambridge University Press. (2009) Page 426.
30. E. Hamid, D. Moraru, J. Tarido, C. S. Miki, T Mizuno, and M. Tabe, Appl. Phys. Lett. **97,** 262101 (2010).
31. A Samanta, D Moraru, T Mizuno, and M Tabe, Sci. Reports **5,** 17377 (2015).
32. A. Afiff, A. Samanta, A. Udhiarto, H. Sudibyo, M. Hori, Y. Ono, M. Tabe, and D. Moraru, Appl. Phys. Express **12**, 085004 (2019).
33. W. Kohn, and J. M. Luttinger, Phys. Rev. **98,** 915 (1955).
34. B. Roche, E. Dupont-Ferrier, B. Voisin, M. Cobian, X. Jehl, R. Wacquez, Phys. Rev. Lett. **108,** 206812 (2012).
35. P.W Anderson. Phys. Rev. **124,** 41 (1961).
36. E Bonet, M. M. Deshmukh, and D. C. Ralph, Phys. Rev. B **65,** 045317 (2002).
37. M. Seo, P. Roulleau, P. Roche, D. C. Glattli, M. Sanquer, X. Jehl, L. Hutin, S. Barraud, and F. D. Parmentier, Phys. Rev. Lett. **121,** 027701 (2018).